\documentstyle[11pt]{article}
 
\def\pa{\partial}
\def\k{\kappa} 
\def\g{\gamma} \def\G{\Gamma}
\def\a{\alpha} 
\def\b{\beta} 
\def\d{\delta} \def\D{\Delta}
\def\e{\epsilon} 
 
\def\k{\kappa}
\def\l{\lambda} 
\def\m{\mu} 
\def\n{\nu}

\def\r{\rho} 
\def\s{\sigma} 
\def\t{\tau}

\def\be{\begin{equation}}
\def\ee{\end{equation}}
     
\begin{document}

\begin{center}
{\large\bf The Immortal Bel--Robinson Tensor}

S. Deser\\
Department of Physics, Brandeis University,\\
Waltham, MA 02454, USA
\end{center}

\begin{quotation}
We present some generalizations, and novel properties,
of the Bel--Robinson tensor, in the context of constructing
local invariants in D=11 supergravity.
\end{quotation}

\noindent{\bf 1. Introduction}

It is a double pleasure to dedicate this lecture to Luis Bel
and to have presented it before both creators of the
Bel--Robinson (BR) tensor \cite{001}.  Indeed, I feel
that I have been one of the prime beneficiaries of this 
amazing quantity over the years \cite{002}, long
after they had moved on the better things.  What is
especially interesting is that many of the applications
of BR have been far from the arena of D=4 general
relativity for which it was created and intended, and 
that it has risen from its original incarnation as a
would-be tensorial energy-density to its avatar as a basic 
member of gravitational supersymmetric multiplets and
invariants; indeed it is there that it takes on 
precisely the stress
tensor role!  Here, I will illustrate the latest twist in
the story of BR, namely its essential contribution in 
the construction of supersymmetric
(SUSY) local invariants for D=11 supergravity. This work
has been performed jointly with D. Seminara; I refer you
to a just-completed compressed
version of our results \cite{004}, to be followed by a
more detailed one.  On the BR side, we
have been joined by J.\ Franklin for some computer-based
calculations; those results are
also to appear in due course \cite{new4}. I refer to those
papers for details.  Here I will only be able to present
the general ideas.

Before getting into BR, let me state the reasons 
why D=11 supergravity and existence of its invariants 
in particular, are once 
again important. There are two distinct motivations: 
(1) if suitable invariants can be constructed,
this will imply that it is a nonrenormalizable local field 
theory, despite its otherwise many attractive
properties and (2) the detailed 
form of these invariants is a
clue to the structure of the M-theory that underlies both 
it and more generally  (D=10) 
string theory as well. 

Since this is a mixed
audience, I will first provide a mini-resume of the relevant
aspects of D=11 supergravity; then we will outline how
SUSY invariants can be systematically constructed from this
action, and in particular the desired lowest order one,
containing scalars quadratic in BR (quartic in curvature).
In the remainder of the paper, we will illustrate some of
the relevant and useful properties of BR, generalized in
three ways: dimension, field content, and tensorial type.

\noindent{\bf 2. ~D=11 Supergravity}

We begin with a brief reminder of this ``uniquely unique"
highest dimensional supergravity theory \cite{005}
whose renewed importance is based on the fact that it is
the field theoretical limit of M-theory, the successor
as well as generalization of D=10 superstring theory.
Recall that  D=11 is the highest dimension in which local
supersymmetry can be achieved
without having to invoke the inconsistent presence of
higher spins than 2
and of more than one graviton.  Its other claims to 
uniqueness are threefold:
(1) there is only one field content permitted, (2) there
can be no ``matter" sources, since no lower spin
supermultiplets exist, and (3) cosmological constant 
extension is forbidden \cite{006}.  There are simply
three fields, the graviton elfbein $e_{\m a}$, the
gravitino $\psi_\m$ and, to balance  the number of
bose/fermi degrees
of freedom, a 3-form potential field $A_{\m\n\a}$ with
totally antisymmetric field strength 
$\pa_{[\b}A_{\m\n\a]} \equiv F_{\m\n\a\b}$, gauge invariant
under $\d A_{\m\n\a} = \pa_{[\a} \xi_{\m\n]}$.
There are then D(D--3)/2 = 44 gravitons,
(D--2)(D--3)(D--4)/3! = 84 form excitations and 
(D--3)$ 2^{[D/2]-1} = 128$ gravitinos.  
The usual SUSY transformation rules
\be
\d e_{\m a} \sim \bar{\a} \G_a \psi_\m \;\;\;\;\;
\d A_{\m\n\a} \sim \bar{\a} \G_{[\m\n}\psi_{\a ]} \; ,
\;\;\;\; \d \psi_\m \sim D_\m \a + (\G F)_\m \a
\ee
leave invariant the action $I_{11}$, whose leading
terms are $(e \equiv \det e_{\m a})$
\be
I_{11} \sim  \int d^{11}x 
\left[ - \textstyle{\frac{1}{4}}
\, \k^{-2} 
eR + \textstyle{\frac{1}{2}}
e\bar{\psi} (\G D+ \G F)\psi -
\textstyle{\frac{1}{48}} eF^2 + 
2\k /(144)^2 \textstyle{\e}^{1..11}
F_{1...}F_{5...} A_{..11}
 \right] \; .
\ee
We have dropped all $\psi^4$ terms; throughout,
$\G$ represents the  appropriate member of the D=11
Clifford gamma algebra. Note the last,
Chern--Simons (CS), term in its initial physics 
appearance, and the explicit presence in CS of the
Einstein constant, whose dimension is 
$\k \sim [L]^{-9/2}$.

What we are after is apparently both quite simple and 
far removed from the BR arena, namely the construction of
SUSY invariants, like $I_{11}$ itself, under the
transformations (1).  This turns out to be neither
simple nor BR-less!  The difficulty lies in the absence
of any superfield methods for D=11, so that it is 
almost impossible to guess the forms of such invariants,
or even to verify candidates if they are guessed.  This
is in contrast with say D=4, where life is much easier and
indeed, where BR takes a natural place in the multiplets
known there from which invariants are then constructible;
as mentioned, BR's role is to replace that of
matter stress tensors that underlie lower spin (global
SUSY) multiplets.

\noindent{\bf 3. Constructing D=11 Invariants}

In this section, we will indicate how it is possible, 
in a straightforward (though lengthy) way to obtain
guaranteed invariants, or more precisely truncations of
such invariants to their leading order in 
$\k h_{\m\n} \equiv g_{\m\n}-\eta_{\m\n}$.  We will also
(for simplicity) stick with their purely 
bosonic on-shell components.  Let me
state the idea: Since $I_{11}$ is an invariant, all
scattering amplitudes (at each loop order) 
it generates also are; 
since SUSY transformations preserve particle number 
(also to lowest order)  we can just ask for the 4-point tree
amplitude with all legs on shell.  This object will only
be a global SUSY invariant, and of course not fully
diffeomorphism invariant -- that would require dressing the
amplitude with infinitely many external graviton legs
as well.  But for our physical purposes, which are
twofold, that suffices, at least with one more caveat to 
remove: A free 2 particle$\rightarrow$2 particle
scattering amplitude is a nonlocal object in general,
since it has an intermediate boson propagator (think
of matter-matter scattering through an intermediate
graviton, $\sim T^{\m\n} (p) (p-q)^{-2} T_{\m\n}(q)$ in
momentum space).  To get a local invariant is 
actually both simple and proves very useful:
one just multiplies
each term by the product (stu) of the usual Mandelstam
variables: one of them knocks out the denominator
($s^{-1}, \; t^{-1}$ or $u^{-1}$ depending on the
channel) while the remaining 4 derivatives 
($tu$, etc.) turn the external graviton and 3-form
polarizations into curvatures and
field strengths, as required by gauge invariance
and explicitly verified by calculation.

The building-blocks in this process are the 
free particle propagators and cubic 
vertices in $I_{11}$, corrected by the 4-point contact
terms to preserve gauge invariance. At tree level
there is no boson-fermion mixing, {\it i.e.}, no
contribution to the bosonic amplitudes from the
fermions.  Indeed, it is clear that the relevant
bosonic 3-vertices are 
\be
V^g_3 \sim \k h \pa h\pa h, \;\;\;\;
 V^{gF}_3  \sim \k h \: FF, \;\;\;\; V^F_3 \sim \k \e AFF
\ee
along with the contact terms $V^g_4 \sim
\k^2 hh\pa h\pa h$ and
$V^{gF}_4 \sim \k^2 hhFF$, required as usual to 
cancel residual gauge
dependence in the 3-vertex contributions.  
The pure $h$ terms come from expanding the 
curvature scalar in $h$, the mixed $hFF$ ones from the
form's kinetic part and finally the pure $A^3$ comes from CS.
The vertices in turn give rise to 4-point localized
amplitudes of the schematic form:
\be
M^g_4 \sim R^4\; , \;\;\;\;
M^F_4 \sim (\pa F)^4 \; , \;\;\;\;
M^{g^{2}F^{2}}_4 \sim R^2(\pa F)^2 \;, \;\;\;\;
M^{gF^{3}}_4 \sim R(\pa F)^3
\ee
where we have already multiplied by stu to get local
forms and the bosonic part of the total 4-point 
function (that also contains 4-fermion and mixed
2 fermi-2 boson contributions) is the sum of the
terms in (4) and guaranteed to be SUSY invariant, as
explained earlier.  The respective scatterings described
by these $M$'s are: 2 graviton-2 graviton, 
2 form-2 form, ``graviton Compton" off a form and finally
gravitational ``bremsstrahlung" from one of the $A$'s
in the 3-point CS vertex (the CS vertex itself is of
course topological, {\it i.e.}, gravity-independent).
Now as soon as one sees quadratic (let alone quartic)
curvature terms, one ought to think of BR.  The very good
reason for this is of course our experience with 
lower-dimensional SUSY, most spectacularly in the form
of the $R^4$-like SUSY invariants in D=4 that was
used to establish the (3-loop)
nonrenormalizability of supergravity 
there \cite{003}.  In that case, it was possible to guess
the shape of the invariants by analogy with 
``squaring" matter
supermultiplets which always included $T_{\m\n}$; the
correct guess was essentially to replace $T_{\m\n}$ by
$B_{\m\n\a\b}$ and hence to expect to have 4-point
SUSY invariants that began quartic in the curvature, 
namely $\sim B^2_{\m\n\a\b}$.  But at D=11, there is on 
the one hand no matter ``crutch" and on the other,
an embarrassment of riches as regards BR; indeed 
we will see that there is
a 3-parameter condidate family to replace the unique totally
symmetric traceless conserved BR we know and love 
in D=4.  There is also another complication here:
unlike the N=1 in D=4, one has the form field in addition to 
the graviton (this is more like so-called $N>1$ models
\cite{007}); the ambition now becomes to have a ``master"
BR that includes the $F$ as well so as to have a simple
elegant form to the overall invariant.  
Fortunately, we have found some general theorems about
gravitational couplings that tell us BR will remain a
basic ``current" also in the present context.
Now recall our two aims in this investigation.  The
first was to decide on whether D=11 supergravity is
nonrenormalizable or finite (since $\k^2$ is dimensional,
it is one or the other -- it cannot be renormalizable),
by exhibiting a local candidate counterterm or showing
none exists.  For this purpose, it is the invariant's
existence, not the details or elegance of its form,
that counts.\footnote{Establishing the nonvanishing
of its coefficient at the relevant loop order where it can
appear requires a separate calculation.  For the present 
model, this was essentially carried out in \cite{new9}.}
But there is another aim that is really
more fundamental, namely to use the D=11 model as a probe
of the as yet unknown underlying M-theory.  For, just
as the nonlocal D=10 superstring theory reduces to
D=10 supergravity in the zero-slope limit, but further 
induces (an infinite series of) corrections,
so should M-theory reduce to D=11. As an amusing sidelight,
these ``zero slope" corrections in the graviton-graviton
sector turn out to have the same form in both D=10
and D=11; they are in turn equivalent to the ``localized"
tree-level pure Einstein 4-graviton scattering amplitude, 
which can be written as \cite{013} 
$t_8 t_8 RRRR$ where $t_8$ is effectively 
an 8-index constant tensor made
out of Lorentz metrics.  
The relation with (BR)$^2$ can then be
obtained by expanding both in a basis
of quartic invariants \cite{011}.  Because of the 
numerous probes of 
M-theory that are being undertaken using
brane dynamics, it is paramount to have exact forms and 
relative coefficients of the corrections (our invariant
being the first such) to be matched against brane
calculations. Clearly,
any underlying unified symmetry (or even good 
``notation" such as BR!)
to be found in the invariant would be very important
for this purpose. Rather than give details of our 
procedure or its results
\cite{004}, I move now to our main subject of BR, 
its properties in arbitrary
dimensions, its link to other invariants, and also
to a solution of an old problem in
\cite{008} attempting to relate BR to gravitational
pseudotensors.

As a concluding remark in this
section, it may be helpful to mention an 
aspect of SUSY and of supergravity 
that sometimes confuses classical relativists.  
Let us phrase it as a question: why should 
purely gravitational quantities (here tree level
graviton scattering amplitudes) that
also happen to arise in supergravity
display any ``SUSY" properties, particularly since
no fermions at all are involved at tree level.
The point is that the {\it supersymmetrizability} that
is inherent in the Einstein action (but {\it not},
say, in its cosmological extension in D=11 as we
have mentioned) already provides strong {\it a priori} 
constraints on its internal properties, quite apart
from whether we choose to implement the SUSY
extension.  Historically, for example, the positive
energy theorem was obtained this way, as was the simple
structure of the D=4 graviton-graviton scattering
amplitude, the result of an
otherwise horribly complicated purely gravitational
calculation \cite{013}.  
The simplicity follows from an underlying helicity
conservation, which is a spinoff inherent to 
supersymmetrizability \cite{009}.

\noindent{\bf 4. BR}

Let's first pay homage to the original, D=4, definition
\cite{001},
\be
B_{\m\n\a\b} = R^\s~_{\m\t\a} \:
R_{\s\n}~^\t ~_\b + ~
^*\!\! R^\s~_{\m\t\a} \: ^*\! R_{\s\n}~^\t~_\b \; ,
\;\;\; ^*\! R^{\m\n}~_{\a\b} \equiv \textstyle{\frac{1}{2}}
\e^{\m\n\l\s}R_{\l\s\a\b} \; .
\ee
This tensor is fully symmetric, traceless, covariantly
conserved on shell ({\it i.e.}, in
Ricci-flat geometries), vanishes {\it iff}
Riemann does and even has
positive energy density $B_{0000}$, just like its model, the
Maxwell stress tensor.  On the other hand, 
charges made from $B$ don't really exist (last paper in 
\cite{002}); it is
a ``zilch" as in fact was (indirectly) 
established much earlier: adding Ricci-dependent terms
converts BR to an identically conserved
quantity \cite{010}, hence without dynamical content.  
However, this in no way
diminishes the importance of BR, in particular in the 
supergravity arena.

First, while still in D=4, we exhibit the promised solution of the
MTW problem: can one simply relate BR to the energy (necessarily
{\it pseudo-}) tensors $t_{\m\n}$ of gravity?  Apart from
the amusing aspect of the question, there is actually a
deep point that is essential to the generalization of BR
to include matter and our form fields
as well as gravity.  Basically, just
by dimensions, BR has two more derivatives (and indices)
than does any $t_{\m\n}$.  One therefore expects any 
such relation to be
of the schematic form $B_{\m\n\a\b} \sim \pa^2_{\a\b}
t_{\m\n}$, though of course it could only hold in some gauge
since neither $t$ nor its derivatives are tensors.
The obvious gauge is that of 
Riemann normal coordinates (RNC) at a point so that all the 
affinities $\G^\l_{\rho\s}$ vanish for simplicity, and we
don't have to worry about ``spreading out" the $\pa^2_{\a\b}$
over the $\G\G$ of $t_{\m\n}$.  It turns out, surprisingly,
that such a relation actually holds without any remainder
(unlike the one in \cite{008});  I omit the 
gory details \cite{new4}.  The result is that, in RNC,
\be
B_{\m\n\a\b} = \pa^2_{\a\b}
(t^{LL}_{\m\n} + \textstyle{\frac{1}{2}} \:
t^E_{\m\n} ) + 0 \; .
\ee
in terms of the Landau--Lifschitz and Einstein 
pseudo-tensors, in a standard normalization.

Some of the simplicity of $B$ in D=4
is, alas, very specific to this (degenerate) dimension,
one in which the number of independent quartic
curvature invariants drops from 7 to 2.  Of particular physical
interest in supergravity there is that its
(unique) square $B^2_{\m\n\a\b}$ can be
written in several equivalent ways,
\be
2 \: B^2 = [(R_{\m\n\a\b})^2]^2 -
[R_{\m\n\a\b} R^{\m\n\l\s}]^2 \sim E^2_4 - P^2_4 
\equiv (E_4 + P_4)(E_4 - P_4)
\ee
where $(E_4, \: P_4)$ are respectively the Euler 
$(^*\! R^*\! R)$ and
Pontryagin $(R^*\! R)$ topological densities of D=4.  
The first equality is already surprising: a square is
also a difference of squares.  The second equality
says it is also the difference of two scalars 
$(E^2_4 , \; P^2_4)$ that can in fact
be chosen to span the quartic invariant basis.  
The last equality,
while obvious, expresses the totally helicity conserving
(or violating, depending on in/out conventions)
character of the 4-graviton scattering amplitude in D=4 
that I referred to earlier \cite{009}; since it is 
essentially proportional to $B^2$. What makes
D=4 special is the quadratic identity
\be
S_{\m\n} \equiv R_{\m\a\b\g} R_\n~^{\a\b\g} -
\textstyle{\frac{1}{4}} g_{\m\n} R^2_{\a\b\g\d}
\equiv 0, \;\;\; D=4
\ee
which is derivable from the fact that any antisymmetrization
over $D+1$ indices vanishes identically at $D$, {\it i.e.},
$R^{\m\n}_{[\m\n} R^{\a\b}_{\a\b} X^\l_{\l]} \equiv 0$
for any X.  Note that the tensor in (8) no longer zero,
but is on-shell
conserved in any D, thereby providing one ``free"
family of BR-like tensors there (it is proportional to the 
trace of (5) in fact).
Next, let me turn to the generalizations 
of $B_{\m\n\a\b}$ in arbitrary dimension $D$.
Beyond D=4, there is actually a 3-parameter
family of what may legitimately be called the successors
of BR, depending on which of its properties one wishes
to keep, apart from that of conservation on (at least)
two indices, say $(\m\n)$.  Of course, one representative of
B in any D is just (5) itself with the $\e\e$ expanded
out,
\be
B_{\m\n\a\b} = R^\s~_\m~^\t~_\a \: R_{\s\n\t\b} 
+ R^\s~_\m~^\t~_\b \; R_{\s\n\t\a} -
\textstyle{\frac{1}{4}} \: g_{\m\n} R_\a~^{\s\t\d}
R_{\b\s\t\d} \; .
\ee
The leading (trace-independent) terms are common to this
whole family. Various choices of the parameters
in this extended BR
will yield specific properties such as tracelessness,
total symmetry, or conservation on all indices.  As a
first step, define the form
$$
\overline{B}_{\mu\nu\alpha\beta}=
B_{\mu\nu\alpha\beta}-\textstyle{\frac{1}{4}} g_{\alpha\beta}
B_{\mu\nu\rho\sigma}g^{\rho\sigma} \eqno{(10{\rm a})}
$$
which enjoys the all-index conservation property.
In terms of $\overline{B}$, the three-parameter
family then reads
$$
B^{(a)}_{\m\n\a\b} \equiv
\overline{B}_{\mu\nu\alpha\beta}+a_1
\overline{B}_{\mu\alpha\nu\beta}+
a_2 \overline{B}_{\mu\beta\nu\alpha}+
a_3 g_{\alpha\beta}\overline{B}_{\mu\nu\rho\sigma} 
g^{\rho\sigma} \; . \eqno{(10{\rm b})}
$$
Again we stress that (only) in D=4, (10b) reduces
to the original, unique, form (5).

Coming now to matter extensions, let me first show that
they naturally arise within the context of (massless)
matter interaction with gravity, even apart from SUSY.
Consider, as a pedagogical example, scattering of massless
scalars through their gravitational couplings; to lowest
order,
\renewcommand{\theequation}{\arabic{equation}}
\setcounter{equation}{10}
\begin{eqnarray}
L_{\rm int} & = & \k h^{\m\n} \; T_{\m\n}(\phi ) \;\;\;\;
T_{\m\n} (\phi ) \; \equiv \; \phi_\m \phi_\n - 
\textstyle{\frac{1}{2}} \eta_{\m\n} \phi^2_\a \nonumber \\
\k h_{\m\n} & \equiv & g_{\m\n}-\eta_{\m\n} \; , \;\; 
\phi_\m \; \equiv \; \pa_\m \phi
\end{eqnarray}
plus the contact term $\sim hh\pa \phi \pa \phi$ needed to
preserve gauge invariance.  Then the resulting 4-scalar
amplitude due to graviton exchange according to (11), using
the graviton propagator $\sim k^{-2}$ and extracting the local
part of this by multiplication with stu, 
is\footnote{I have been informed that Teyssandier and 
Bel (unpublished) constructed
similar $B(\phi )$ ; a Maxwell
$B_{\m\n\a\b}(F_{\l\s})$ has been defined in \cite{012},
following the $\pa_{\a\b} t_{\m\n}$ gravitational idea.}
\be
B_{\m\n\a\b} (\phi ) = \phi_{\a\m} \phi_{\b\n} +
\phi_{\b} \phi_{\a\n} - \eta_{\m\n}
\phi_{\a\s} \phi_\b~^\s
\ee
where $\phi_{\m\n} \equiv \pa^2_{\m\n} \phi$.  This quantity
is separately $(\m\n )$ and $(\a\b )$ symmetric and 
conserved on $(\m \n )$ on-shell $(\Box \phi = 0)$.  It
can even be covariantly completed, to become covariantly
conserved (on Einstein + scalar shell), by turning all
$\pa_\m \rightarrow D_\m$ and adding to (12) 
curvature-dependence of the form
\be
2 \D B_{\m\n\a\b} (\phi ) \sim R_{\m\s\n\b} \;
\phi^\s\phi_\a + (\m\n ) + (\a\b ) \; ,
\ee
{\it i.e.}, adding the 3 terms indicated by the symmetrizations.

So our interest is not so much to play BR for its own sake.
Rather it is to see how we are led, within the rather 
different framework of SUSY invariant scattering amplitudes,
not only to this sort of $B_{\m\n\a\b}(F_{\g\d\s\t})$ for
our form fields, but also to unified
\be
B_{tot} \equiv B(g) + B(F)
\ee
expressions.  As might be expected from the universal vertex
coupling forms, the $B(F)$ should look very similar.
Indeed, $B(F)$ (with again admits of several parameter
extensions) has the leading
$\pa^2_{\a\b} T_{\m\n} (F)$ form, namely
\be
B(F)_{\m\n\a\b} = 
\pa_\a F_\m \pa_\b F_\n +
\pa_\b F_\m \pa_\a F_\n -
\textstyle{\frac{1}{4}} \, \eta_{\m\n}
\pa_\a F\pa_\b F , \;\;\;
\pa^\m B(F)_{\m\n\a\b} = 0
\ee
where all omitted indices are traced in an obvious way.
Its conservation holds on shell, $\pa^\m F_{\m\n\a\b} = 0$,
using the Bianchi identities $\pa_{[\l}F_{\m\n\a\b]} \equiv 0$.
One can even go further and incorporate fermionic $B(\psi_\l )$
where $\psi_\l$ is the gravitino potential.  It has a form
(again $\sim \pa^2_{\a\b} T_{\m\n} (\psi )$) similar to
the gravitational $B$, and is quadratic in the fermionic
curvature $f_{\m\n} = \pa_\m\psi_\n - \pa_\n\psi_\m ,\;
B(\psi ) \sim \k^2 (\bar{f} \G \pa f)$.  Its form is relevant
to the full SUSY invariant that includes the 4-fermion
and mixed, 2-fermion---2-boson, amplitudes.

{\bf 5. Conclusions}

After 40 years, 
not only has the original unique D=4 gravitational BR of
\cite{001} led us from (5) to the higher complexities 
(and usefulness!) of expressions such as
(14), but also to a different class of ``currents" that
have proven to be equally essential building-blocks of
SUSY invariant expressions.  These include gravitational
4-forms 
\be
P_{\m\n\a\b} = \textstyle{\frac{1}{4}}
R^{\l\s}_{[\m\n} R_{\a\b]\l\s}
\ee
that are closed rather than conserved; they are of course
easily overlooked in D=4 where they reduce
essentially to the Pontryagin scalar 
$P_{\m\n\a\b} \rightarrow \e_{\m\n\a\b} (R^*\! R)$.  
Then there are mixed gravity-form conserved currents 
\be
C^{RF}_{\m\n\r ;\a\b} \equiv
\pa_\l  \left[ R^\a~_{(\a}~^{[\l}~_{\b )} F_\s~^{\m\n\r ]}
\right]
-\textstyle{\frac{1}{6}} R^\s~_{(\a}~^\l ~ _{\b )}
\pa_\l F_\s~^{\m\n\r} \; ,
\ee
that involve both bosonic
building blocks of D=11 supergravity and the list goes on.

This tribute to BR and its serendipitous
importance in supergravities at all dimensions, as well as
the many ways in which it generalizes, should
assure us that BR industries (like its founders)
still have a great future.

I thank my collaborators J. Franklin, and especially
D. Seminara.  This work was supported by NSF grant
PHY93--18511.

\end{document}